# Covalent-bonding-induced strong phonon scattering in the atomically thin WSe$_2$ layer

*Young-Gwan Choi$^{†\#}$, Do-Gyeom Jeong$^†$, H. I. Ju$^†$, C. J. Roh$^†$, Geonhwa Kim$^†$, Bongjin Simon Mun$^†$, Tae Yun Kim$^§$, Sang-Woo Kim$^§$, and J. S. Lee$^{†*}$*

$^†$Department of Physics and Photon Science, School of Physics and Chemistry, Gwangju Institute of Science and Technology, Gwangju 61005, South Korea

$^§$School of Materials Science and Engineering, Sungkyunkwan University, Suwon 16419, South Korea

In nano-device applications using 2D van der Waals materials, a heat dissipation through nano-scale interfaces can be a critical issue for optimizing device performances. By using a time-domain thermoreflectance measurement technique, we examine a cross-plane thermal transport through mono-layered (n=1) and bi-layered (n=2) WSe$_2$ flakes which are sandwiched by top metal layers of Al, Au, and Ti and the bottom Al$_2$O$_3$ substrate. In these nanoscale structures with hetero- and homo-junctions, we observe that the thermal boundary resistance (TBR) is significantly enhanced as the number of WSe$_2$ layers increases. In particular, as the metal is changed from Al, to Au, and to Ti, we find an interesting trend of TBR depending on the WSe$_2$ thickness; when referenced to TBR for a system without WSe$_2$, TBR for n=1 decreases, but that for n=2 increases. This result clearly demonstrates that the stronger bonding for Ti leads to a better thermal conduction between the metal and the WSe$_2$ layer, but in return gives rise to a large mismatch in the phonon density of states between the first and second WSe$_2$ layers so that the WSe$_2$-WSe$_2$ interface becomes a major thermal resistance for n=2. By using photoemission spectroscopy and optical second harmonic generation technique, we confirm that the metallization induces a change in the valence state of W-ions, and also recovers a non-centrosymmetry for the bi-layered WSe$_2$.

With an advent of exfoliation technique for van der Waals (vdW) materials, there have been tremendous progresses in 2D nanotechnology using graphene, transition metal dichalcogenides (TMD), and so on[1-4]. Novel functionality can be realized in artificial heterostructures prepared as a multi-stacking of different layers exploiting a relatively weak vdW bonding strength[5-7]. Since such artificial structures can be made ultimately in a form of ultrathin atomic layers, one of the key issues in those devices should be a heat management[8-10]. The cross-plane thermal resistance in ultrathin 2D materials should be contributed to mainly by the boundary (Kapitza) resistance due to their infinitesimal thickness. In understanding the heat dissipation through 2D materials, there have been several previous works about the nano-scale heat transport based on the phonon transmission spectrum across the interfaces. Chen *et al.* used a non-equilibrium molecular dynamics (NEMD) simulation and atomistic Green's function (AGF) approach to clarify the thermal transport in the metal/graphene/metal geometry[11]. They argued that depending on a metal-graphene bonding strength, a significant mismatch in the phonon density of states (pDOS) can occur between two adjacent graphene layers, and the graphene interlayer can be a major resistance source. In addition, Yan *et al*. focused on the thermal transport through the metal/TMD/metal junction with an AGF method[12]. They found that a strong bonding between the metal and TMD leads to an electron injection from metal to TMD, and the cross-plane thermal transport can be modulated as a result of the change in the pDOS. Ong *et al*. demonstrated that an encasement of graphene with a top $SiO_2$ layer introduces a new heat transport channel for the low frequency phonons, and hence the graphene-substrate Kapitza resistance can be significantly reduced[13].

Besides such theoretical works, there have been a few other experimental studies on the heat transport through the van der Waals materials, such as graphene and $MoS_2$[14-16]. Nevertheless, much less experimental reports have been made yet particularly about the heat transport through the interface between metal-vdW materials. In this work, we investigate the thermal transport through ultrathin $WSe_2$ layers in contact with metal layers of Al, Au, and Ti by using time-domain thermoreflectance (TDTR) technique which, as a pump-probe optical technique (Fig. 1(a)), provides us with key information about the cross-plane thermal boundary resistance. In particular, we focus on the role of metal contacts for minimizing the heat barrier in the 2D devices using vdW materials. Whereas all three metal layers lead to a change in the electronic structure of $WSe_2$ to have a finite density of states at the Fermi level, such tendency becomes stronger in the order of Al, Au, and Ti[14,15]. Therefore, we expect to learn how the electronic bonding strength between metal and $WSe_2$ layers influences on the phonon transport through ultrathin vdW materials. Using the optical second harmonic generation technique, we found a clear signature of a non-centrosymmetric crystal structure of otherwise centrosymmetric bi-layer $WSe_2$ due to the metallization of the top $WSe_2$ layer. Also, we observed a corresponding change in the valence state of W-ions using x-ray photoemission spectroscopy. In the case of the thermal transport behavior, we observed a large difference of the thermal boundary resistance depending on the metal film. In particular, we could demonstrate that when $WSe_2$ is more strongly bonded to the metal, e.g., Ti, the thermal transport through the metal-$WSe_2$ interface becomes better, but in return it becomes worse for the $WSe_2$-$WSe_2$ interface, as schematically shown in Fig. 1(b). To minimize the thermal resistance in the 2D devices using vdW materials, it is therefore required to consider not only the hetero-interface formed between the metal and vdW layer, but also the homo-interface formed between two adjacent vdW layers.

**Distinct thermal properties of the mono- and bi-layered $WSe_2$.**

We prepared mono- and bi-layer WSe$_2$ flakes, termed $n$=1 and $n$=2, respectively, on an Al$_2$O$_3$ (001) substrate. Optical images of the samples, shown in Figs. 2(a)-(c), indicate that each flake has a triangular shape with a lateral size of about 10–20 μm. Mono- and bi-layered flakes are clearly distinguished from the color contrast; the brighter (darker) regions correspond to the bi-layer (mono-layer) flakes. To examine how the thermal transport across ultrathin WSe$_2$ flakes is influenced by a metal layer in a direct contact with WSe$_2$, we prepare three kinds of metal (M), i.e., Al, Au, and Ti. Thickness of the metal layer is 100 nm for Al, 20 nm for Au, and 2 nm for Ti. An additional 100 nm thick Al layer is prepared on top of Au and Ti.

We first examine a spatial distribution of the cross-plane thermal property by mapping a TDTR signal. (Refer to the experimental section for the details of the measurements.) Figures 2(d)-(f) display spatial mapping results of the -$V_{in}$/$V_{out}$ signal for M=Al, Au, and Ti, respectively, where the signal at each point is taken with a 5 μm spatial resolution and a 1 μm interval. Ratio value, -$V_{in}$/$V_{out}$, is picked up at 1 ns of the time delay, and it is roughly proportional to thermal boundary conductance. In this TDTR mapping data, mono- and bi-flakes are clearly distinguished as in the optical images taken for the same area. This suggests that the thermal boundary resistance should be given differently for $n$=1 and $n$=2 although their thicknesses are different only by about 0.6 nm. Furthermore, it is worthwhile to note that the color contrast between $n$=1 and $n$=2 is the largest for the Ti contact. Giovannetti *et al*. and Kang *et al*. suggested that graphene and TMDs can be metallized due to the covalent bonding with the metal layer[14,23]. Moreover, Chen *et al*. and Yan *et al*. suggested that these additional electron doping can perturb the pDOS of graphene and TMDs[11-12]. In this respect, the electron injection to the upper WSe$_2$ layer will have a strong influence on the thermal transport in the metal-TMD-insulator heterostructure, and such effect should appear differently depending on the bonding strength of the metal layer in contact with TMD materials.

**Modulation of the thermal boundary resistance by a metal-WSe$_2$ bonding.**

For a more quantitative analysis of the thermal boundary resistance, we obtained a full decay profile of $-V_{in}/V_{out}$ signals for each region of the samples, i.e., $n=1$ and $n=2$ as well as $n=0$ where the metal has a direct contact with the Al$_2$O$_3$ substrate. As displayed in Figs. 3(a-c), the three areas exhibit distinct time-dependent behaviors of the signal, and the relative values of each region are given consistently with the mapping results shown in Fig. 2. By considering the heat diffusion equation, we could model these time-dependent signals; the entire system is assumed to be composed of two layers of the metal transducer and the sapphire substrate, and their interface in between. In particular, the mono- and bi-flakes are so thin that their contribution to the thermal transport is taken into account as a part of the thermal boundary resistance. Each flake was measured three times at different positions, and 20 different flakes were examined in total. An average of all these results are taken with their standard deviations as error bars. The thermal conductivity of the sapphire is determined as 28.7±2.5 W/mK. Details of the modelling and analysis can be found elsewhere[20].

Figures 3(d-f) display the obtained thermal bound resistance (TBR) R of $n=0$, 1, and 2 for M=Al, Au, and Ti, respectively. For $n=0$, TBR at the metal-substrate interface is in the range of ~1–2×10$^{-8}$ m$^2$K/W. Although TBR can be varied depending on a surface condition and an evaporation technique, these values are in good agreement with previous reports of the TBR for the similar interfaces, such as 2.2×10$^{-8}$ m$^2$K/W for Au/Al$_2$O$_3$ and 0.95×10$^{-8}$ m$^2$K/W for Al/Al$_2$O$_3$[21]. For all metal contacts, there is a large increase of R by about 2–6×10$^{-8}$m$^2$K/W as the WSe$_2$ flakes are inserted between the metal and the substrate. Note that a similar increase of TBR was experimentally observed when the graphene or MoS$_2$ layer is inserted between Au/Ti and SiO$_2$[14,15],

or when the graphene is located between Al and Cu[16]. As the number of the WSe$_2$ layer increases from $n=1$ to $n=2$, a further increase of R is observed. Such an increase of R is particularly large for the Ti contact.

To get a better insight into these behaviors of the boundary thermal resistance, we consider a bonding density at the metal-WSe$_2$ junction. As the covalent bonding between the metal and WSe$_2$ becomes stronger, the electron density at their interface becomes higher. Hence, we can consider this bonding density as a measure of the bonding strength between the metal and WSe$_2$. We take the corresponding values from the previous theoretical works by Liu *et al*. and Kang *et al*., i.e., 0.008/Å$^3$ for Al, 0.018/Å$^3$ for Au, and 0.029/Å$^3$ for Ti[17,18]. Here, we introduce the ratio of thermal boundary conductance G (=1/R) to denote a phonon transmittivity. Figure 4(a) shows that as the bonding density increases, $G(n=1)/G(n=0)$ increases, but $G(n=2)/G(n=1)$ seems to decrease.

Here, the relative thermal conductance between $n=1$ and $n=0$ is less than 0.4 for all three cases, and such a reduction in $G(n=1)$ is attributed to the van der Waals bonding nature of the WSe$_2$ layer or the interface roughness[15]. Nevertheless, the relative value of $G(n=1)$ is the largest for the Ti contact where the bonding density is at the maximum. Given the perfect crystalline quality of the interface, the thermal boundary resistance arises as a result of the scattering of the phonons which is due to the pDOS mismatch between the metal and WSe$_2$ layer[14]. As the metal-2D layer bonding becomes stronger, the pDOS of the 2D layer shifts toward that of the bonding material[11,12], and hence the pDOS mismatch can be reduced leading to the enhancement of $G(n=1)$. In the same context, we can discuss the possible role of the WSe$_2$/WSe$_2$ interface in the cross-plane thermal transport by comparing $G(n=2)/G(n=1)$. Contrary to $G(n=1)/G(n=0)$, $G(n=2)/G(n=1)$ exhibits the smallest value for the Ti contact with the largest bonding density. Although the strong chemical bonding between Ti and WSe$_2$ layers has a significant effect on the pDOS of WSe$_2$, such effect would be limited only up to the first layer of WSe$_2$. This leads to the significant pDOS mismatch between upper and lower WSe$_2$ layers, and hence results in the stronger phonon scattering as schematically shown in Fig. 1(b)[11].

To highlight the thermal boundary resistance occurring at the WSe$_2$-WSe$_2$ interface, we examine the additional resistance for *n*=2 compared to the case of *n*=1. Namely, we obtain ΔR= R(*n*=2)-R(*n*=1), and find that it ranges from $1.5 \times 10^{-8}$ m$^2$K/W for Al to $4 \times 10^{-8}$ m$^2$K/W for Ti as shown in Fig. 4(b). These values are unexpectedly large when they are considered as boundary resistances at the homo-junction formed by the same kind of materials. Actually, the metal/semiconductor heterojunctions have the TBR values in a range of $0.5 - 10 \times 10^{-8}$ m$^2$K/W[22], which can be confirmed also for *n*=0 as shown in Fig. 3(b). WSe$_2$ double layers are successively deposited by using a chemical vapor deposition technique, and hence we can ignore the effect of interstitial voids, bubbles, or dirt that should be considered importantly in other sample preparation techniques, such as exfoliation and transfer methods[23-25]. Rather, since ΔR increases linearly with an increase of the bonding density, we consider a strong hybridization between the metal and the first WSe$_2$ layer as a major reason of such a large ΔR. Actually, similar behaviors can be found also in the graphene layers in contact with the metal. In the tri-layer graphene sandwiched by metal layers, Chen *et al*. showed using the NEMD simulation that the TBR at the graphene-graphene interface becomes larger as the metal-graphene bonding becomes stronger. We include such behavior with red squares in Fig. 4(b); a temperature-drop (ΔT) across the graphene-graphene interfaces increases in proportion to the bonding strength[11]. These results of WSe$_2$ and graphene strongly suggest that the interface formed by two vdW materials could be a major thermal resistance in ultrathin devices made of vdW layers particularly in contact with metal layers.

**Electronic and crystal structural properties of surface and interface of WSe$_2$.**

To confirm the bonding configuration between the metal and TMD materials, we conduct x-ray photo-emission spectroscopy (XPS) measurement. Among three metal contacts, we check the Al case. Figure 5(a,b) displays W 4f peaks obtained for a bare WSe$_2$ and the Al(5 nm)/WSe$_2$ specimen. Two peaks around 32.2 eV and 34.4 eV correspond to the doublet W 4f$_{7/2}$ and W 4f$_{5/2}$, respectively[26-28], of the WSe$_2$ composition. The other two peaks around 35.8 eV and 38.0 eV correspond to the doublet W 6f$_{7/2}$ and W 6f$_{5/2}$, which are attributed to the WO$_3$ precursor[26]. We assume that doublet W 6f$_{7/2}$ and W 6f$_{5/2}$ responses do not change depending on specimens due to the constant density of WO$_3$ seeds. With this assumption, comparison between the two set of XPS responses reveals that the doublet W 4f$_{7/2}$ and 4f$_{5/2}$ responses are reduced when WSe$_2$ is in contact with Al. Actually, this is attributed to the formation of W$_2$Se$_3$, which originates from the covalent bonding between Al and WSe$_2$ layers[26-28].

A similar bonding nature between the metal contact and WSe$_2$ layer can be manifested by utilizing the optical second harmonic generation (SHG) measurement[29]. For the mono-layer WSe$_2$, it shows a strong SHG signal due to the broken inversion symmetry; a three-fold rotational symmetry is clearly observed for the SHG signal obtained as a function of the sample azimuth in a normal-incidence reflection geometry. And, this anisotropy is in agreement with the 3*m* point group[30-32]. For bi-layer TMDs, on the other hand, the centrosymmetry is usually restored, and the SHG intensity is negligible as observed for the bare bi-layer WSe$_2$ (Fig. 5(c))[30-33]. Interestingly, the bi-layer WSe$_2$ in contact with Al gives a strong SHG signal; the SHG intensity amounts to almost half of the intensity observed for the mono-layer WSe$_2$. This observation reflects a change in a structural symmetry, and can be understood in a similar way with the electronic state change observed by XPS; the metal-WSe$_2$ bonding induces a change in the structural property of the upper

WSe$_2$ layer, and its combination with the intact lower WSe$_2$ layer leaves the bi-layer WSe$_2$ to be inversion asymmetric[30-33].

**Discussion.**

Consequently, we can consider the changes in electronic state and crystal symmetry of the top WSe$_2$ layers as one of the key factors in understanding the observed significant modulation of the thermal boundary resistance in the metal/WSe$_2$/sapphire structure. Namely, the metallization of the WSe$_2$ layer actually occurs by a strong bonding between metal and WSe$_2$ which is expected to modify the phonon density of states in the WSe$_2$ layer. Although this process would increase the phonon transmission across the metal-WSe$_2$ interface, it can have an opposite effect on otherwise the homo-interface between adjacent WSe$_2$ layers. To realize 2D devices with an efficient heat dissipation property, the net thermal transport should be optimized by considering the heat dissipation efficiencies not only of hetero-junctions, e.g., the metal-WSe$_2$ interface, but also of the homo-junctions, e.g., the WSe$_2$-WSe$_2$ interface.

**Sample preparation and characterization.**

The monolayer and bilayer WSe$_2$ were synthesized by chemical vapor deposition (CVD) method. CVD chamber is divided into two zones. The Se (0.2 g, 99.99 %) powders were put on an alumina boat placed in the middle of the first heating zone, and the WO$_3$ powders (0.01 g, 99.995 %) were put on an alumina boat placed in the middle of the second heating zone. The sapphire substrates for deposition and growth of WSe$_2$ were put at the right side of WO$_3$ powders in the second heating zone. Argon gas for carrier gas and hydrogen for reductant were used with flow rate of 190 sccm and 10 sccm, respectively. During the synthesis of monolayer (and bilayer) WSe$_2$, first and second zones were heated up to 150 °C and 925 °C (and 875 °C for bilayer $WSe_2$), respectively, for 37 min and then maintained for 30 min. After finishing the growth, the chamber was naturally cooled down to room temperature. Using the multi-target e-beam evaporator, 100 nm thick Al is deposited for all three cases, whereas Au and Ti are pre-deposited onto a $WSe_2$ specimen with a thickness 10 nm and 2 nm, respectively. The thickness of the entire metal layer is measured by acoustic echo signals from TDTR measurements[34-35].

**Time-domain thermoreflectance (TDTR) measurement.**

We used TDTR to observe temporal temperature excursion and also to extract thermal parameters, such as the thermal conductivity and thermal boundary resistance[14,20,35-36]. As a light source for this pump-probe measurement, we used femtosecond laser pulses with a 785 nm wavelength and a repetition rate of 80 MHz. The beam was split for the pump and probe sources by a polarizing beam splitter, and the pump beam was modulated by an electro-optic modulator with a frequency 10 MHz. As schematically shown in Fig. 1(a), the pump-induced temperature change of the Al thermal transducer was monitored by the reflectivity change of the probe beam of which arrival time at the sample position was adjusted by a 300-mm-long delay stage. Both beams were focused by a 10× magnitude objective lens onto the sample where the beam size was about 5 $\mu$m of $1/e^2$ diameter. The beam size and its circular shape were maintained in an entire range of the time-delay[37]. To avoid an unnecessary detection of the pump beam, we separated pump and probe beams in a wavelength domain by adopting a two-tint method, and used appropriate optical filters: 785 nm long pass filter for the pump beam and 785 nm short pass filters for the probe beam[38]. A probe signal was detected by a photo-diode, and fed into a fast pre-amplifier and an electrical 10 MHz band-pass filter. Finally, the signal was taken by using a radio frequency lock-in amplifier which provided in-phase ($V_{in}$) and out-of-phase ($V_{out}$) signals. We monitored their ratio, i.e., $V_{in}/V_{out}$ with a variation of the sample position as well as the pump-probe delay time[20].

**X-ray photoemission spectroscopy.**

The configuration of a chemical bonding between metal and TMD material was examined by utilizing x-ray photoemission spectroscopy. Entire measurements were carried out under an ultra-high vacuum condition at room temperature. The excitation source of x-ray was an aluminum K-alpha with an energy 1486.6 eV, and a hemispherical electron analyzer from ScientaOminkron, Hipp-3 analyzer, was employed.

**Optical second-harmonic generation.**

We employed an optical second-harmonic generation technique to examine a structural symmetry of the mono- and bi-layered $WSe_2$ flakes. We adopted a femtosecond laser with an 800 nm wavelength and an 80 MHz repetition rate as a fundamental wave source, and focused it on the flakes with a beam size of 1 μm in a normal incidence. The second harmonic wave was collected in a reflection geometry by using a photo-multiplier tube. Polarization states of the fundamental and the second harmonic waves were controlled by using a half wave plate and polarizer, respectively.

AUTHOR INFORMATION

**Corresponding Author**

*E-mail: jsl@gist.ac.kr

**Present Addresses**

[#]Department of Energy Science, Sungkyunkwan University, Suwon 16419, South Korea.

**Author Contributions**

Y.-G.C. and J.S.L. designed the experiments together with D.-G.J. and H.I.J. T.Y.K. and S.-W.K. fabricated the bare $WSe_2$ samples, and Y.-G.C. and H.I.J. deposited metal films on them. Y.-G.C., D.-G.J. and H.I.J. performed time-domain thermoreflectance measurements. C.J.R. and Y.-G.C. conducted the second-harmonic generation measurement. G.K. and B.S.M. conducted x-ray photoemission measurements. All authors discussed the results. Y.-G.C and J.S.L wrote the manuscript. J.S.L supervised the project.

**Notes**

There is no competing interest.

ACKNOWLEDGMENT

This work was supported in part by the Science Research Center and the Basic Science Research Program through the National Research Foundation of Korea (NRF) funded by the Ministry of Science, ICT & Future Planning (Nos. 2015R1A5A1009962, 2018R1A2B2005331). This work was supported by the "GRI (GIST Research Institute)" Project through a grant provided by GIST in 2018.

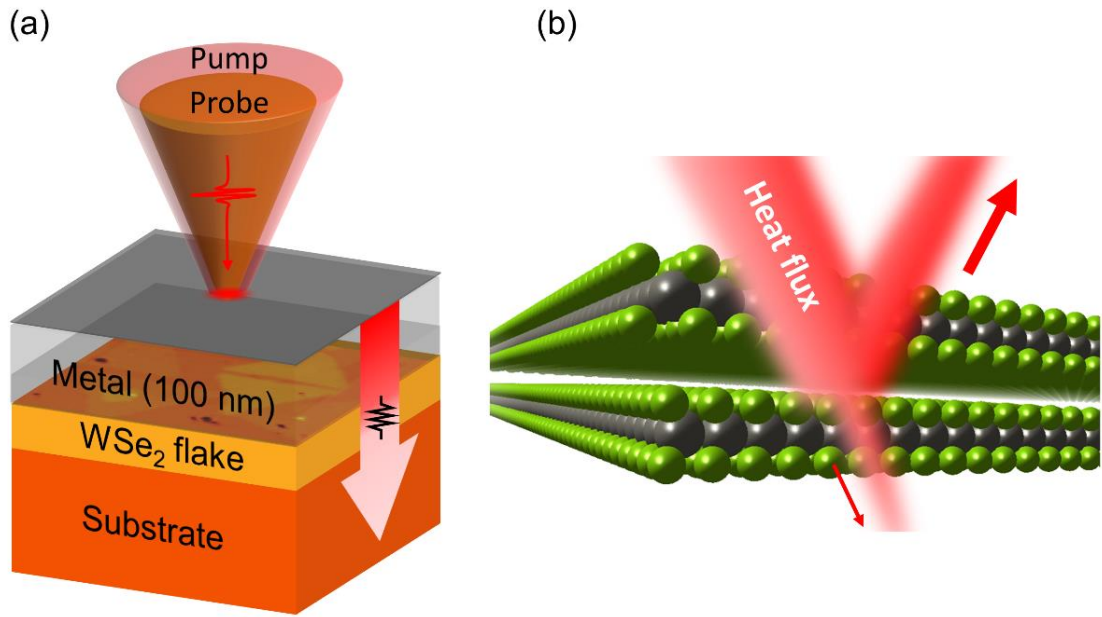

**Figure 1.** Schematic pictures of (a) the sample part in time-domain thermoreflectance experiment and (b) a cross-sectional view of a heat flux through $WSe_2$ layers. In (a), an optical pump beam heats the metal surface, and the heat becomes dissipated through $WSe_2$ flakes and finally to the substrate. Probe beam monitors a temperature change of the metal surface in terms of the reflectivity change. Figure (b) highlights a large thermal resistance of the $WSe_2$-$WSe_2$ interface which is due to a significant mismatch in the phonon density of states of neighboring $WSe_2$ layers.

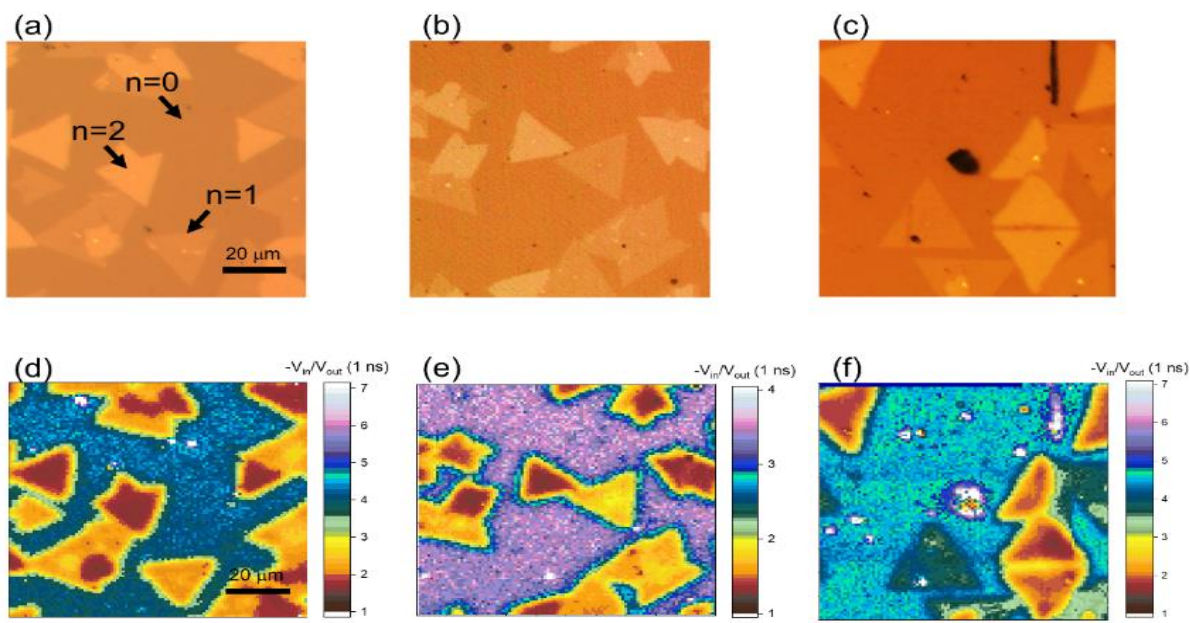

**Figure 2.** (a-c) Optical images of WSe$_2$ flakes prepared on an Al$_2$O$_3$ substrate. (d-f) Time-domain thermoreflectance mapping results for Al/WSe$_2$/Al$_2$O$_3$, Al/Au/WSe$_2$/Al$_2$O$_3$, and Al/Ti/WSe$_2$/Al$_2$O$_3$. The ratio signal -V$_{in}$/V$_{out}$ is monitored at the pump-probe time delay fixed at 1 ns. In both images, mono- and bi-flakes have distinctive contrasts.

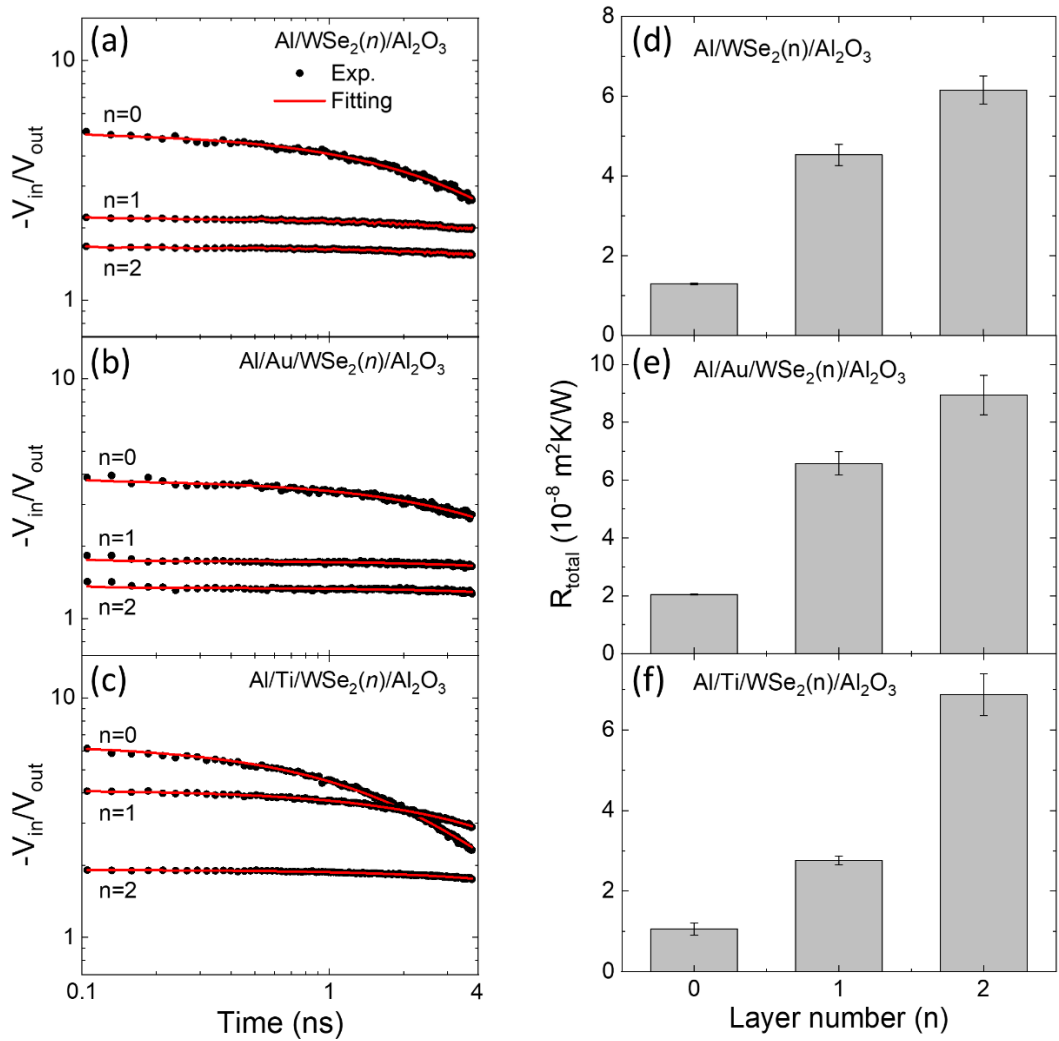

**Figure 3.** (a-c) Time-domain thermoreflectance data (symbol) for Al/WSe$_2$/Al$_2$O$_3$, Al/Au/WSe$_2$/Al$_2$O$_3$, and Al/Ti/WSe$_2$/Al$_2$O$_3$. In each case, the number of WSe$_2$ layers is indicated by $n$=0, 1, and 2. The fitting curves (solid lines) for the time-domain data are also shown. (d-f) Extracted thermal boundary resistance (TBR) values depending on the layer number $n$.

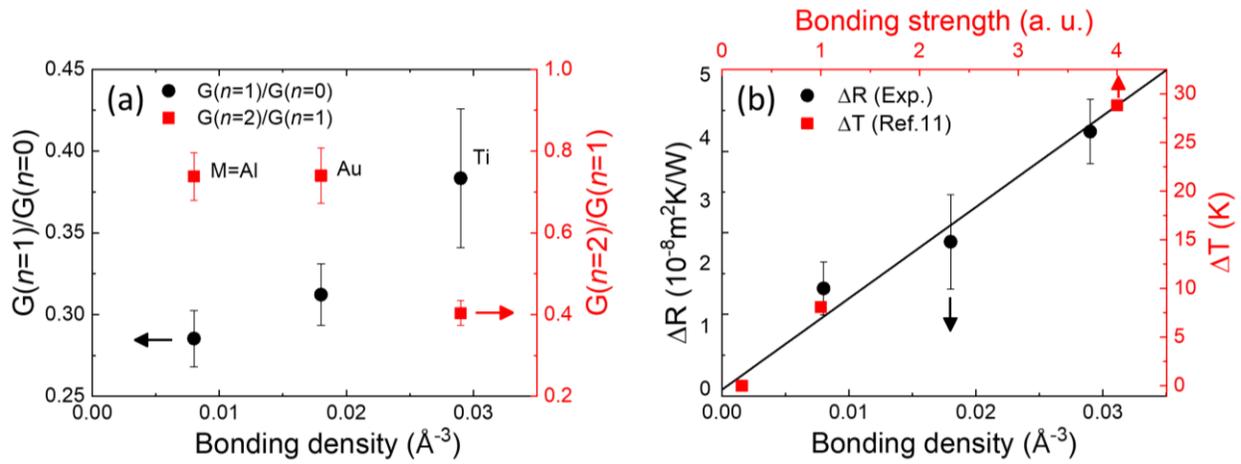

**Figure 4.** (a) Thermal boundary conductance ratio between $n$=1 and 0 (left axis), and between $n$=2 and 1 (right axis). The values are displayed as a function of the bonding electron density at the metal-WSe$_2$ interface which are quoted from Refs. 17,18. (b) Thermal boundary resistance difference between $n$=1 and 0 (black circle) and a temperature-drop in multi-layer graphene (red square) quoted from Ref. 11.

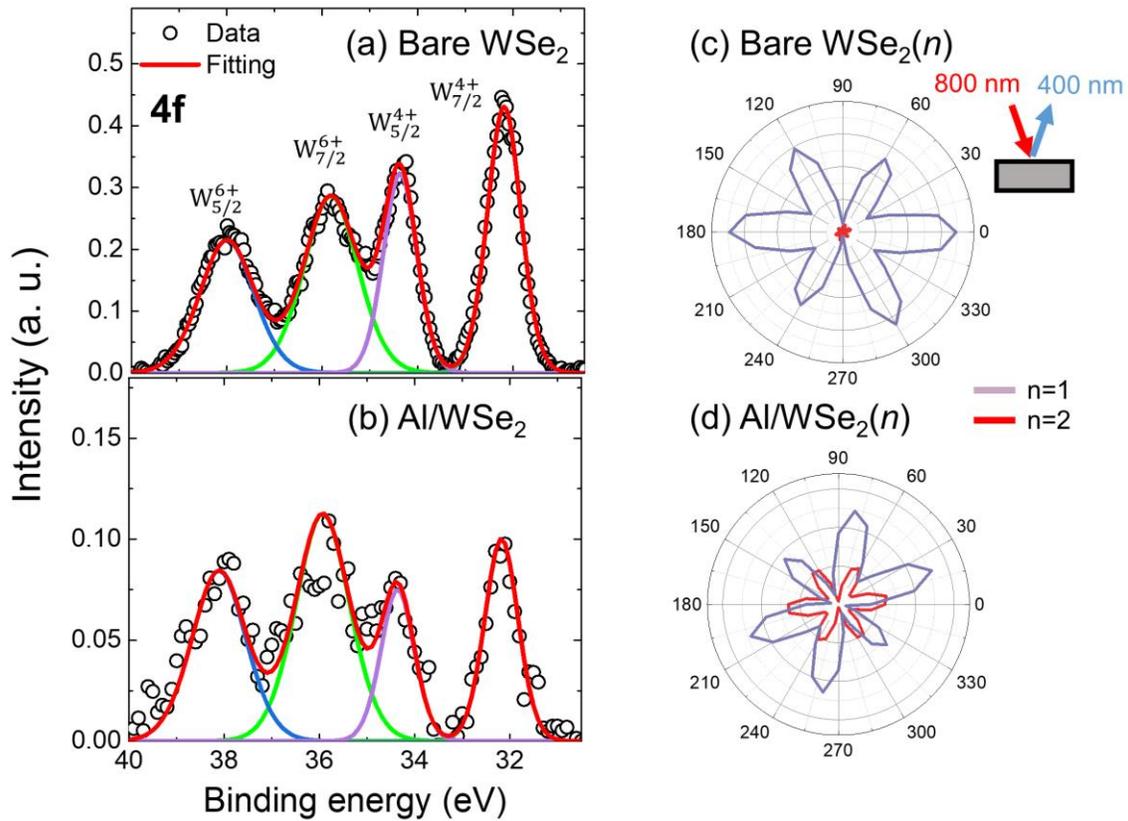

**Figure 5.** (a,b) X-ray photoemission spectroscopy data around the W 4f peaks for a bare $WSe_2$ (a) and $Al/WSe_2$ (b). Four peaks can be fitted with contributions of $W^{6+}$ and $W^{4+}$ states, and two peaks of $W^{4+}$ states become weaker for $Al/WSe_2$ sample. (c,d) Azimuth-dependent second-harmonic generation (SHG) intensity from $WSe_2$ flakes with a layer number $n$=1 and 2. Inset shows a schematic of the SHG experiment where red and blue arrows indicate the fundamental (800 nm) and frequency-doubled (400 nm) waves, respectively. The SHG intensity for $n$=2 is ignorable for a bare $WSe_2$, but it becomes almost half of the $n$=1 case for a $WSe_2$ with an Al layer on its top.